\pdfoutput=1
\documentclass[aps,twocolumn,showpacs,groupedaddress,\if{lengthcheck}\fi,
amsmath,amssymb,floatfix
]{revtex4}

\usepackage{epsf,
amsmath,graphicx,amssymb,subfigure}

\newcommand{\bs}{\mbox{\boldmath{$\sigma$}}}
\newcommand{\bS}{\mbox{\boldmath{$\Sigma$}}}
\begin{document}
\title{Modeling the Relaxation of Polymer Glasses under Shear and Elongational Loads}
\author{S. M. Fielding,$^1$ R. L. Moorcroft,$^1$  R. G. Larson,$^2$ M. E. Cates$^{3}$}
\affiliation{$^1$ Department of Physics, Durham University, Science Laboratories, South Road, Durham DH1 3LE, UK \\
$^2$ Department of Chemical Engineering, University of Michigan, Ann Arbor, MI 48109-2136, USA\\
$^3$ SUPA, School of Physics and Astronomy, The University of
  Edinburgh, JCMB Kings Buildings, Edinburgh, EH9 3JZ, UK}
\date{\today}
\begin{abstract}
{ 
Glassy polymers show ``strain hardening'': at constant extensional load, their flow first accelerates, then arrests. Recent experiments under such loading have found this to be accompanied by a striking dip in the segmental relaxation time.  This can be explained by a minimal nonfactorable model combining flow-induced melting of a glass with the buildup of stress carried by strained polymers. Within this model, liquefaction of segmental motion permits strong flow that creates polymer-borne stress, slowing the deformation enough for the segmental (or solvent) modes to then re-vitrify. Here we present new results for the corresponding behavior under step-stress shear loading, to which very similar physics applies. To explain the unloading behavior in the extensional case requires introduction of a `crinkle factor' describing a rapid loss of segmental ordering. We discuss in more detail here the physics of this, which we argue involves non-entropic contributions to the polymer stress, and which might lead to some important differences between shear and elongation. We also discuss some fundamental and possibly testable issues concerning the physical meaning of entropic elasticity in vitrified polymers. Finally we present new results for the startup of steady shear flow, addressing the possible role of transient shear banding.  
}
\end{abstract}

\maketitle

\section{Introduction} Understanding the flow of polymeric materials is a central issue in their manufacture and performance. For molten systems, profound insights into polymer rheology can be obtained by combining simple ideas on entropic elasticity (stress arising from entropy changes among random-walk chains) with conceptually simple but mathematically sophisticated ideas on how molten polymer chains can move under strongly entangled conditions \cite{McLeishReview,LarsonBook}. Despite their ubiquity in products of all kinds, progress in understanding the flow of polymer glasses has been much slower. This is perhaps not surprising given that even `simple' glasses, arising in hard-sphere atomic or colloidal materials with no polymeric degrees of freedom, lead to constitutive models of roughly similar complexity to that of molten polymers \cite{BraderPRL2,BraderPRE}. These non-polymeric glasses  \cite{Molecular,Metallic,Colloidal}
share with polymeric ones \cite{Ediger,Struik} the feature of undergoing slow plastic deformation in response to applied stress. This applies at least in the temperature range just below the glass transition temperature $T_g$ which we address here. (At much lower temperatures, brittle fracture may occur.) However in general they do not show strain-hardening as seen in polymers: simple glasses are melted by stress, and at constant load then remain indefinitely in a flowing state \cite{BraderPNAS}.

Faced with the complexity of the polymer glass problem, a number of modeling strategies are possible. A crucial question is how much glass physics and how much polymer physics to try to include. One important avenue has been to first develop a largely empirical approach to the glass sector, coupled initially with minimalist polymers (modelled as elastic dumb-bells, for example) and then add more polymeric detail such as additional Rouse modes. Exemplary of this approach is the EGP (Eindhoven Glassy Polymer) model \cite{EGP,EGP3}, whose antecedents date back to the work of Haward and Thackray \cite{EGP2}.  

On the other hand, our understanding of flow in simple glasses has improved greatly in recent years, through advances in microscopic \cite{BraderPRL2,BraderPNAS} and mesoscopic \cite{SGR,KS1,STZ} theory, much of it well tested experimentally. Crucial to glass rheology (but not yet under good control in the microscopic approaches \cite{BraderPRL2,BraderPRE}) is physical aging: a quiescent glass becomes more solid over time, but `rejuvenates' under flow into a liquefied state. This physics is distilled by so-called `fluidity' models, which feature a time evolution equation for a single structural relaxation rate, called the fluidity \cite{Fluidity,Fluidity2}. In the so-called `simple aging' scenario, the inverse of the fluidity, which is the structural relaxation time $\tau$, for a quiescent glass increases linearly with its age \cite{SGR,Aging,Struik}. A slow steady flow cuts off this growth at the inverse flow rate; increasing the shear rate (e.g., in startup of steady shear) causes rejuvenation \cite{Fluidity,Fluidity2}.

We believe that this improved understanding of the glass sector justifies a renewed approach to the problem of polymer glasses, again starting initially by coupling a suitable glass model to the simplest (dumb-bell) description of the polymer degrees of freedom. (Clearly, a future objective is to improve the polymeric part of the description, but it is first instructive to see what physics can emerge at dumb-bell level.) This renewed approach was initiated by three of us in \cite{PRL}, mainly within the specific context of explaining strain hardening under tensile loading, and its effects on segmental relaxation rates, as observed in \cite{Ediger}. In the present paper, we give a fuller presentation of the same approach, and also give new predictions for the corresponding creep-test experiments  in a simple shear geometry. To the best of our knowledge such tests have not yet been done, at least for the experimental system addressed in \cite{Ediger}, so this work (which uses only the fit parameters from the extensional data in that paper) offers new and separately testable predictions of our simple approach. 
We also present some additional results on shear startup focussing on the possible role of transient shear banding.

\section{Strain Hardening}
\begin{figure}
\begin{center}
\includegraphics[width=72mm]{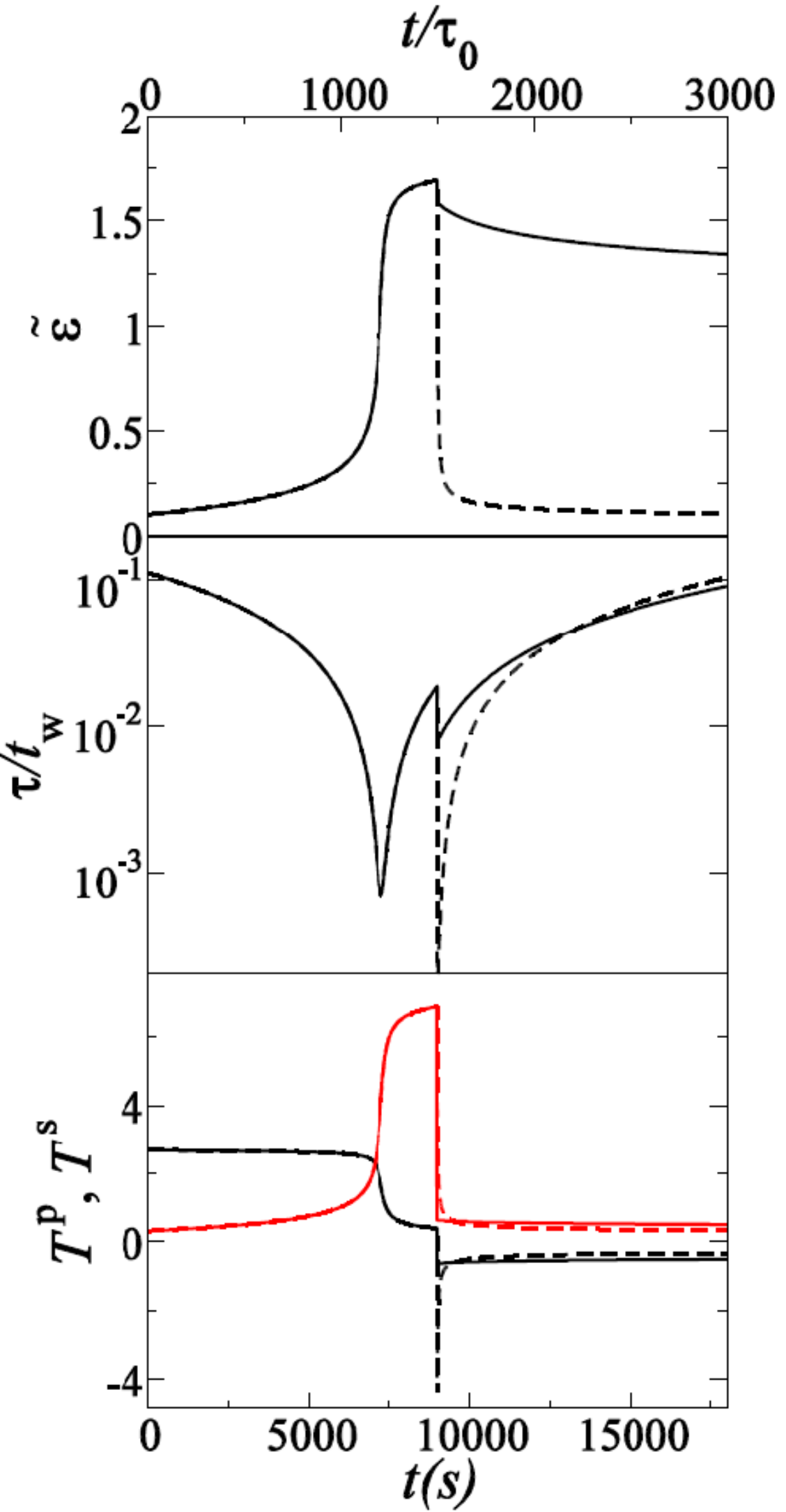}
\caption{(Color online) Solid curves: local strain
$\tilde{\varepsilon}=\exp\varepsilon-1$ \cite{Ediger}, reduced relaxation time $\tau(t)/t_w$ 
and tensile
stresses $T^{p,s}=G^{p,s}(\sigma_{zz}^{p,s}-\sigma_{xx}^{p,s})$ of the
polymer (p) and solvent (s) during loading of an infinite uniform
cylinder. Here $t_w = \tau(0^-)$ is the age of the system when the experiment begins. Parameters are $G^s/G^p =8.5$, $\mu=12.5$, $t_w/\tau_0=10^4$, $\tau_0=6$s; applied force / initial area $f=2.7G_{\rm p}$. 
(The curve for $T^p$, in red, initially lies below $T^s$ but crosses it during strain hardening.) 
The unload results for
the basic model ($\theta=1$) is shown dashed; the solid curve
after unload has $\theta = 0.1$. The horizontal axis is marked both in dimensionless model units (top) and real time (converted using $\tau_0$), bottom. (As explained in \cite{PRL}, the numerical solver introduces, in lieu of inertia, a small additional fluid viscosity $\eta_n = 0.05 G^p\tau_0$ into Eq.\ref{zero}, whose magnitude has negligible influence on these plots.)
\label{fig_one}}
\end{center}
\end{figure}

In polymeric glasses, the interplay between polymeric and glassy degrees of freedom causes new properties to emerge. Particularly striking among these is the response to loading of the segmental relaxation time $\tau(t)$, which controls the rate of local rearrangements. Lee et al \cite{Ediger,Ediger2} found that $\tau(t)$ falls steadily during the early stages of elongational deformation, and then falls more sharply, reaching a small fraction $\sim 10^{-3.3}$  
of its initial level before dramatically rising again. (Fig.\ref{fig_one} shows all these features for a theoretical curve that was presented in \cite{PRL} as the best fit to the data of Lee et al; the unloading part is discussed separately below.)
The rise in $\tau(t)$ happens as the local strain rate starts to drop: the latter property defines onset of the so-called strain-hardening regime. 

Various elements of this scenario have been confirmed in coarse-grained and molecular simulations \cite{Simulations1,Simulations2,Simulations3,Simulations4}, but we believe ours is the first simple theoretical picture of it to capture the key features just described. For instance the theories of Chen and Schweizer \cite{KS1,KS2,KS3,KS6}, which emphasize the role of stress-induced hopping over barriers (following Eyring \cite{Eyring}), seem unable to account for the striking dip in segmental relaxation time. 
More generally the assumption of a stress-dependent fluidity sits uneasily with the idea of aging in glasses, which causes time dependence of the fluidity even under constant stress \cite{Aging,SGR,Struik}. 
In our own work, in common with more fundamental theories of simple glasses \cite{BraderPRL2,BraderPRE}, fluidity arises primarily by strain-induced barrier crossing (deformation forces cages to break) as opposed to stress-induced effects. This distinction means that $\tau(t)$ depends on the accumulated flow history rather than solely on the present state of stress; the latter possibility is clearly ruled out by the data of \cite{Ediger}. 

Previous work to incorporate aging and flow-rejuvenation into polymer glass theory led to the Eindhoven Glassy Polymer model \cite{EGP}, whose viscosity (or equivalently $\tau(t)$) is controlled by the evolution of a state parameter $S$ that is age- and strain-dependent. Such a state parameter might have various physical meanings and -- especially if one is also interested in the temperature and pressure dependence which is very important for applications \cite{EGP} -- its time evolution is in general likely to be complicated. Given this,  it is hardly a criticism of that work that empirical simplifying assumptions
were made concerning the time evolution of $S$. 

\begin{figure}
\begin{center}
\includegraphics[width=76mm]{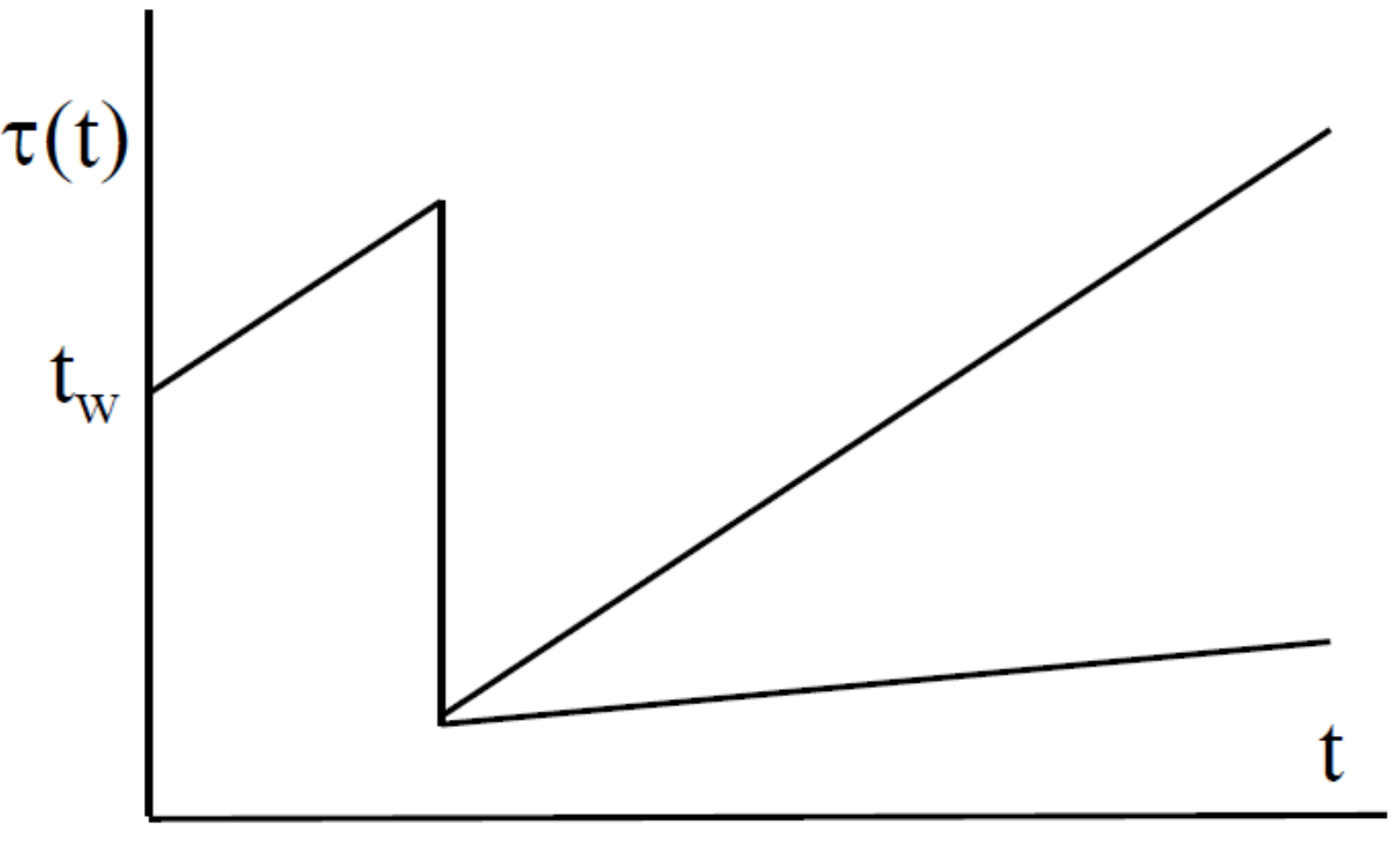}
\caption{Schematic evolution of the relaxation time $\tau(t)$ in a sample with $\tau(0^-) = t_w$ -- which is its age or ``waiting time" in our model -- subjected later to a step strain causing a sudden drop in $\tau$. In the simple aging picture, $\tau(t)$ rebuilds from this point with the same slope as before (upper curve). However, if aging and rejuvenation are factorable, the slope of the curve drops by the same factor as $\tau$ does (lower curve).}
\label{fig_two}
\end{center}
\end{figure}

Crucially, however, among these assumptions is one that cannot be reconciled with the aging and rejuvenation scenario encapsulated by the fluidity approach. Specifically, the EGP model assumes aging and rejuvenation to have {\em factorable} effects on $S$. 
(The authors of \cite{EGP} already acknowledged the potential limitations of this approximation.)
Thus,  strain-induced rejuvenation causes a (cumulative) loss of structure, encoded in a reduction in $S$, which is remembered indefinitely and reduces all subsequent relaxation times by the same factor. This contrasts with fluidity models, where strain-induced fluidization resets the clock for aging but does not change its rate. Fig.\ref{fig_two} schematically compares these two cases following a step strain.

Although we have not seen any direct attempts in the literature to fit the data of \cite{Ediger} to the EGP model, it seems qualitatively clear from this discussion that the EGP model is unlikely to account for the rapid recovery in $\tau(t)$ on entering the strain hardening regime. However it should be possible to do this with relatively simple fluidity-inspired modifications to the EGP model, and indeed the work begun in \cite{PRL} and continued in this paper could be interpreted as a step in exactly that direction. Meanwhile though, the EGP's precepts as currently formulated are arguably unsuited to the regime of strong fluidization arising just below $T_g$ that was addressed experimentally in \cite{Ediger} and in recent glass rheology theories \cite{BraderPRL2,BraderPNAS,SGR,Fluidity,Fluidity2}. This regime may have different physics from the more strongly solidified regime further below the glass transition (though it is not clear even then that factorable behavior would result).
The latter regime is of course highly relevant to in-use mechanical deformation of glassy polymers (in car fenders and the like) but the near-$T_g$ regime is certainly also important in molding and other processing steps for such materials. Possibly both regimes could be spanned by a single, generalized evolution equation for $S$ (or $\tau$) but we leave this aspect to future work. 

\section{Dumb-Bell Fluidity Model}
\label{short}
Despite several recent efforts \cite{KS1,KS2,KS6,hoy}, creating a comprehensive theory of rheological aging in polymer glasses remains a formidable task. 
Here we recall the ingredients of a minimal model \cite{PRL}, combining just two key elements of any such theory (nonfactorable aging/rejuvenation, and the strain dependence of polymer-borne stresses), which semiquantitatively explains many of the results reported in \cite{Ediger}. Subsequently (Sections \ref{more},\ref{mmore}) we give a more detailed discussion of the underlying physical assumptions than was possible in \cite{PRL}. 

We consider polymeric dumb-bells \cite{LarsonBook} suspended in a glassy fluid or `solvent' whose microscopic relaxation time $\tau(t)$ obeys a fluidity-type equation showing simple aging and flow-rejuvenation. The glassy fluid could be a genuine solvent, or alternatively, if no actual solvent is present, it could describe the short-scale, relatively fast degrees of freedom that control the local dynamics of segments. The model distinguishes these from the slower degrees of freedom of large sections of chain, represented here for simplicity by the added dumb-bells. This `division of labor' is well established in polymer melt theory where, for instance, the Rouse model adopts a local friction (equivalent to a time-independent segmental relaxation time $\tau$) that can equally be provided by drag against a solvent or against segments of other chains. Indeed there is little difference in behavior between molten polymers and sufficiently concentrated polymer solutions \cite{LarsonBook}.
One interesting question is whether this also applies to polymer glasses: are there important differences in behavior between a single-component polymer glass, and a concentrated polymer solution in a molecular solvent below its $T_g$? Our model assumes that any such differences are controlled primarily by the relative magnitudes of the elastic moduli of polymeric and solvent-like degrees of freedom (see below). Dilution of the polymer component would delay the onset of strain hardening and enhance the solvent yield behavior that precedes this. However, we do not pursue this issue in the current paper.

For simplicity we treat the dumb-bells first as linearly elastic elements -- as is valid for the entropic elasticity arising at modest deformations in the molten state \cite{LarsonBook}. However, in Sections \ref{more},\ref{mmore} we discuss further the true nature of the polymer stress in polymeric glasses which is not solely entropic in character \cite{KS6,Robbins}, and show how this can be partially incorporated into the model.

We start by defining a deviatoric stress tensor 
\begin{equation}
\bS = G^p(\bs^p-{\bf I}) + G^s(\bs^s-{\bf I}) \label{zero}
\end{equation} 
where $\bs^p$ and $\bs^s$ are dimensionless conformation tensors for polymer and solvent, $G^{p,s}$ associated elastic moduli (see below), and ${\bf I}$ the unit tensor. For entropic dumb-bells of concentration $n$ and equilibrium mean square end-to-end distance $\langle R^2\rangle_0$, $G^{p} = 3nk_BT$ and $\bs^p = 3\langle{\bf RR}\rangle/\langle R^2\rangle_0$. The dumb-bell conformation tensor $\bs^p$ is taken to obey a standard upper-convected Maxwell model, which can be derived \cite{LarsonBook} by considering the dynamics of an ensemble of dumb-bells whose endpoints are (i) advected by flow (ii) subject to a linear spring force inwards along the line connecting them and (iii) subject to independent Brownian diffusion:
\begin{equation}
\dot\bs^p+{\bf v}.\nabla\bs^p = \bs^p.\nabla{\bf v} + (\nabla{\bf v})^T.\bs^p - \alpha(\bs^p - {\bf I})/\tau \label{one}
\end{equation}
Here the structural relaxation time $\tau^p=\tau/\alpha$ is proportional to, but much larger than, that of the solvent, which we denote $\tau$. In the simplest models of dense, molten, but unentangled polymers, $\alpha=N^{-2}$ with $N$ the polymerization index \cite{LarsonBook}. On the other hand in a lightly crosslinked elastomeric network, as studied experimentally in \cite{Ediger}, one expects $\alpha = 0$.
Note that so long as $\alpha$ is small enough, its precise value has very little effect on our numerical calculations, at least for the experimental protocol of \cite{Ediger}. This feature is explained in Section \ref{load}.

We now turn to the solvent stress $G^s\bs^s$. Bearing in mind that (unless a true solvent is present) this stress is also polymeric in origin, we take the relevant conformation tensor to obey another upper-convected Maxwell equation, now with relaxation time $\tau$:
\begin{equation}
\dot\bs^s+{\bf v}.\nabla\bs^s = \bs^s.\nabla{\bf v} + (\nabla{\bf v})^T.\bs^s - (\bs^s - {\bf I})/\tau \label{two}
\end{equation} 
Because local degrees of freedom outnumber the chain-scale ones we expect $G^s > G^p$.

In a departure from molten polymer models, in which the Maxwell time is a fixed quantity, we next assume this structural relaxation time to have its own dynamics, governed by a fluidity equation
\begin{equation}
\dot\tau +{\bf v}.\nabla\tau = 1 - (\tau-\tau_0)\lambda 
\label{three}
\end{equation}
Here we have defined the following scalar invariant measure of flow rate
\begin{equation}
\lambda({\bf D}) \equiv  \mu\sqrt{2\hbox{\rm Tr}({\bf D}.{\bf D})} \label{four}
\end{equation}
with ${\bf v}$ the fluid velocity and ${\bf D} = (\nabla{\bf v}+(\nabla{\bf v})^T)/2$. The parameter $\mu$ is an order unity dimensionless quantity which controls how effectively flow prevents aging. This in turn sets the steady-state relaxation time as a function of strain rate via
\begin{equation}
\tau_{ss} = \tau_0 +\frac{1}{\lambda}  \label{foura}
\end{equation}

Without flow, $\tau$ increases linearly in time at a solidification rate $\dot\tau({\bf D = 0})$. This quantity is dimensionless and for simplicity we set it to unity. An alternative (given that direct measurements of the aging of segmental relaxation rates were not made in \cite{Ediger}) would be to leave it in the model as a freely floating parameter. We have refrained from doing this, but would expect other choices (replacing the 1 in Eq.\ref{three} with some other number) to give quantitative rather than qualitative changes. 

With flow present and aging hypothetically switched off, $\tau$ would undergo deformation-induced relaxation towards $\tau_0$ which is a `fully rejuvenated' (microscopic) value. According to Eq.\ref{three} this relaxation occurs at the rate $\lambda$. 
In steady shear, $\lambda = \mu|\dot\gamma|$ and $\tau$ varies inversely with strain rate $\dot\gamma$ in accord with microscopic theory \cite{FuchsCates,BraderPRL2}.
For uniaxial elongation at strain rate $\dot\varepsilon$, (\ref{four}) reduces to  $\lambda = \mu\sqrt{3}|\dot\varepsilon|$.
Note that in this simple fluidity model, the rejuvenation of $\tau$ is essentially strain-induced \cite{BraderPNAS} but, in contrast to the factorable model of \cite{EGP}, can be rapidly reversed by subsequent aging.

Our chosen fluidity model can in principle be embellished with additional parameters, so as to incorporate further relevant information about glass rheology. We do not pursue this in detail here, but mention one interesting example, which is to write instead of Eq.\ref{three}
\begin{equation}
\dot\tau +{\bf v}.\nabla\tau = 1 - r\lambda(\tau-\tau_0) -\frac{(1-r)\tau}{\lambda^{-1}+\tau_0}\label{fourbb}
\end{equation}
For our chosen model, $r=1$, the dynamics of $\tau$ is essentially strain-induced, so that a pulse of fast flow (whose limit defines a `step strain') is just as effective at fluidizing the system as when the same strain is applied more slowly. In contrast, for $r = 0$, step strains have no fluidizing influence at all. (This model of fluidity was explored in \cite{RobynPRL}.) Both limits give the same result, Eq.\ref{foura}, for steady-state flows.
One expects the true behavior of glasses to lie between these two limits --- but in a manner that can itself depend on the strain amplitude, and sign, in relation to the previous deformation history \cite{BraderPRL1,BraderPRL2}.

To complete the specification of our model, we add the standard equations of mass and force balance for an incompressible fluid of negligible inertia:
\begin{eqnarray}
\nabla.{\bf v} &=& 0 \label{fourb}\\
\nabla.[\bS+ 2\eta{\bf D}] &=& 0\label{fourc}
\end{eqnarray}
In (\ref{fourc}) we have added a small additional Newtonian viscosity $\eta$. This is included solely for numerical reasons: it ensures that a state of force balance can always be found without fluid acceleration, so that inertia can be set to zero from the outset. The alternative is to introduce a small inertia; in practice for the flows considered here, this choice (along with the precise value of $\eta$) has no discernible effect on any of the results.

\section{Elongational Loading}
\label{load}
In \cite{PRL} we presented numerical data for uniaxial extension flows within a lubrication approximation appropriate to long cylindrical samples undergoing homogeneous deformation. As demonstrated there, the data replotted again here in Fig.~\ref{fig_one}, for both the `local strain' $\tilde{\varepsilon}(t)=\exp\varepsilon-1$ and the segmental relaxation time $\tau(t)$ closely resembles the experimental results of \cite{Ediger} for a cylindrical sample of polymer glass subject to a sudden tensile loading introduced after a period of aging at rest, up to the point where the load is removed (but not beyond, see Section \ref{more} below).  

Because the sample was lightly cross-linked, $\alpha$ was set negligibly small; the experimental protocol of \cite{Ediger} then specifies the engineering stress $f = 16$MPa (and also the time $t_u = 9400$s at which unload later occurs) and measures directly the initial relaxation time $\tau(0) = 6\times 10^4$s. 
There remain four material parameters in our model: $G^p,G^s,\tau_0$ and $\mu$. Of these, $G^p \simeq 6$MPa was deduced from the deformation in the strain-hardened regime just before unload; $G^s \simeq 50$MPa and $\mu \simeq 12.5$ were in turn estimated from the step-change in $\tau$ during initial loading, and from the separately measured slope \cite{Ediger} of the parametric `effective flow curve' $\dot\varepsilon(\tau)$. Hence the only unconstrained parameter 
in fitting the dip in $\tau(t)$ was $\tau_0$, which was found to obey $\tau_0\simeq 6$s. 

Striking features of the experimental data, reproduced by our simple model, include: the tenfold initial drop in $\tau$ on applying the load; its subsequent further decline to a sharp minimum $\tau_{\rm min}\sim 10^{-3.3}\tau(0)$ near the point of maximum elongation rate; and its rapid rise from that minimum towards a strain-hardened plateau. Other comparisons using the same model parameters are given in \cite{PRL} and also offer encouraging agreement between model and experiment.
 
Fig.~\ref{fig_one} shows (alongside strain and relaxation time) the tensile stresses $T^{p,s}$ carried by polymer and solvent respectively. Quite striking is the transfer of load from solvent to polymer which ushers in the strain hardening regime. 
Indeed, according to our simple model, it is this transfer that explains strain hardening. Within the model, what matters is that the glassy solvent has a certain yield stress $\Sigma^s_Y$,  initially exceeded by the applied load. After an initial step-down in $\tau$ caused by the step strain on loading, the material starts to flow continuously which fluidizes it further. As a result its strain rate accelerates, giving positive feedback so that $\tau(t)$ plunges downwards. As deformation builds up, however, an ever growing share of the applied stress is carried not by the solvent but by the stretching polymer chains, which have little or no relaxation on this timescale. This polymeric stress causes the flow rate to drop: the solvent, whose stress now obeys $\Sigma^s < \Sigma^s_Y$, therefore starts to solidify. It enters a simple aging regime in which the flow progressively grinds to a halt.

It is notable that this explanation (which also directly explains the remarkable behavior seen for $\tau(t)$) does not require any nonlinearity in the polymeric response. (Such nonlinearity might nonetheless be present; see Section \ref{more} below.) It does require the existence of two parallel channels for supporting stress, so that the unloading of the solvent allows its solidification at late times. Accordingly, we predict that strain hardening of this character will never emerge in a simple, one component (molecular or colloidal) glass. 

An interesting consequence of this scenario is that the polymers, whose Maxwell time scales with that of the solvent but is much larger, can never reach a state of terminal relaxation once strain hardening has begun. Even if (departing from the protocol of \cite{Ediger}) the load were never removed, once the polymers are carrying most of the stress, the solvent re-freezes and, in accord with the simple aging picture, the polymer relaxation time starts growing as $\tau^p \sim t/\alpha$. Since $\alpha$ is large, at no stage can the terminal regime ($t \ge \tau^p$) be reached. This is why, as mentioned in Section \ref{short}, the value of $\alpha$ is almost irrelevant to our predictions, so long as it is small enough. Because the terminal zone is not reached, this experimental protocol will not distinguish a cross linked network from unlinked chains of high molecular weight.

As illustrated in Fig.~\ref{fig_two}, models that factorize aging and rejuvenation effects are seriously challenged by the rapid recovery of $\tau$ after the dip. (A multimode spectrum \cite{EGP3} is unlikely to help here.) With simple aging, such factorization predicts $\tau \sim(t+t(0))f(\epsilon)$, so that if the segmental relaxation times falls from its pre-deformation value $t(0)$ to a small value $\tau = ft(0) = \tau_{min}$ at the dip, a tenfold recovery to $\tau \sim 10 \tau_{min}$ does not occur until $t\sim 10t(0)$
which for the experiment of \cite{Ediger} means $t\sim 6\times 10^5$s. This prediction is at least two orders of magnitude longer than the experimental timescale ($\sim 3000$s), and is an underestimate since the dip occurs during a period of rapid flow (so aging is less fast than in a quiescent system). In contrast, for our simple-aging fluidity model, the recovery time is predicted to be of order $t\sim 10 f t(0) = 300$s. This is also an underestimate, for the same reason, and on that basis is in better accord with the data. 

\section{Elongational Unloading}
\label{more}

A significant shortcoming of the model becomes apparent when the sample is unloaded. Here the experiments show a modest drop in $\tau$ immediately on removing the load, followed by a gradual recovery towards the pre-deformation value. The dotted line in Fig.~\ref{fig_one} shows the prediction based on Eqs.~(\ref{one}--\ref{four}); $\tau$ drops, but then falls much further before recovering. (The solid line is a modified model, discussed below, which much more closely resembles the experimental data \cite{PRL}.)  Moreover, for stresses well above the yield stress, this re-solidification only occurs after an almost complete recovery in the strain. 

This means that the model fails to capture a major aspect of polymer glass behavior, which is that the glass can be deformed plastically and will then hold its shape with only modest relaxation once the load is removed. The problem arises because in the strain-hardened regime, the polymers carry a large elastic tensile stress, which generally exceeds $\Sigma^s_Y$. 

Upon unloading, this polymeric stress is unbalanced by the external traction and therefore acts backwards on the vitrified solvent, causing it to yield. The resulting evolution of
$\tau(t)$ resembles a repeat of the initial loading experiment, but run in reverse; only when the polymer stress (now equal and opposite to the solvent stress) falls below $\Sigma^s_Y$ does revitrification of the solvent occur. By this time, for typical parameters, the sample has almost recovered its original shape. 

One possible candidate for the discrepancy with experiment lies in the response of $\tau(t)$ to imposition of step strains, as arise at both the onset and the removal of the extensional load.
As described following Eq.\ref{fourbb}, this can be subtle for glasses (see \cite{VoigtmannSoft}) and one could imagine circumstances in which curtailing the sharp drop in $\tau$ immediately on unloading would have a strong effect on what happens subsequently. For this reason we have numerically explored the effect of reducing $r$ in Eq.\ref{fourbb} on the post-unload behavior. For the basic model ($r=1$) the post-unload drop in $\tau(t)$ is by a factor of order 20 (for the model parameters used previously, see Fig.~\ref{fig_one}); this factor can indeed be reduced by dropping $r$ (the experimental reduction factor is of order 2) in accord with the discussion of step strain response following Eq.\ref{fourbb}. However, this adjustment cannot prevent the near-complete strain reversal following unload, and the argument in the previous paragraph (which does not depend on $r$) explains why. The incomplete strain reversal can only be explained if the polymer stress falls below the solvent yield stress after only a modest degree of reversal has taken place.

Thus we are led to seek a mechanism that can allow the polymer stress to drop much faster on unloading than would normally be the case for elastic dumb-bells. An important clue is that the value of $G^p \simeq 6$ MPa needed to fit the loading data is one order of magnitude larger than the rubbery modulus of the same material above its glass transition (see, e.g., \cite{SHPaper1}). This accords with a widely held literature view that the strain-hardened modulus of polymer glasses does not primarily arise from single-chain entropic elasticity  \cite{KS6,Robbins}. The larger modulus could come from intrinsic nonlinearity (finite extensibility) in the stress-strain curve of a single chain or it could be caused by the buildup of local mechanical interaction forces at chain-chain contacts. These two mechanisms are much more similar than they might first appear. Indeed, for an inextensible rod in an elongational flow field, the stress is primarily a viscous solvent stress and caused by the obligatory distortion of fluid streamlines to maintain a no-slip boundary condition with the rod. Given that we are treating the local friction between chains by pretending each chain is embedded in a glassy solvent, the chain-chain contact forces can likewise be thought of arising within the solvent.

As proposed in \cite{PRL}, our currently preferred explanation
invokes a well established piece of physics that arises when flexible polymer chains are placed in an extensional flow fast enough for them to extend rapidly relative to their own relaxation time. This requires $\dot\varepsilon\tau^p\gg 1$; within our model and for well aged samples, this condition always holds when $\alpha$ is small. For instance, in steady state $\tau^p\sim \mu \dot\varepsilon/\alpha$ so that the fast flow regime applies, independent of the actual strain rate, whenever $\alpha \ll \mu^{-1}\sim 0.1$.

In that regime it is argued that short subsections of chain quickly stretch close to full extension locally, forming a quasi-one-dimensional filament containing hairpin-like kinks \cite{LarsonKink,Hinch}. Further stretching is mediated by migration and annihilation of neighboring kinks of opposite sign. During this process, much of the stress carried by the polymers is not entropic-elastic, but instead caused by viscous drag against the extended subsections of chain (which behave locally as inextensible rods). 

If one tries to describe this situation with a dumb-bell model, then, to match the resulting stress, an enhanced value of $G^p$ must be invoked. However, if the flow is now suddenly stopped (or reversed) much of this extra stress disappears on a relatively rapid timescale \cite{LarsonKink}. This is not set by the global relaxation time of the whole chain, as a dumb-bell picture would assume, but the time it takes for locally straight subsections to crinkle back into an entropic conformation. A similar physical process arises when chains are fully stretched at all scales, again causing a rapid loss of polymer stress on unloading with only partial relaxation of global polymer conformations \cite{CCH}. 

A proper treatment of these effects requires explicit analysis of short-scale polymeric degrees of freedom, corresponding to higher Rouse modes rather than the single mode retained in the dumb-bell description \cite{LarsonBook,LarsonKink}. A phenomenological `renormalized dumb-bell' representation can however be gained by assuming that, on unloading, the effective polymer modulus abruptly drops by a `crinkle factor' $\theta$ so that $G^p\to \theta G^p$. 
Here the lower value should roughly correspond to the standard entropic-elastic modulus \cite{footcon}. 

A numerical value $\theta =0.1$ for the crinkle factor is suggested by the previous observation that $G^p$ is about ten times larger than the entropic modulus observed above $T_g$.  The solid lines in Fig.~\ref{fig_one} show the results of this choice, which are much closer to the experimental behavior both for the strain and for $\tau(t)$. Post-unload, the polymer stress acting backwards on the solvent is now safely below the solvent yield stress; the result is a modest drop and then slow increase in $\tau(t)$. This is accompanied by only partial restoration of the initial shape with arrest in a finite state of deformation when the entropic polymer stress $\theta G^p \bs^p$ falls below the yield threshold for the glassy solvent.

\section{Some Issues of Principle}
\label{mmore}
Given the crudeness of the crinkle-factor picture, this appears a quite satisfactory outcome. The approach does highlight certain questions of principle however. For instance, it is sometimes argued that entropic contributions to thermodynamic quantities such as stresses can only arise under conditions where kinetic exploration of configuration space actually occurs \cite{Ma}. Holders of this view might expect the entropic part of the polymer stress to switch off as the solvent freezes, requiring an extra creep of the sample during freezing until the non-entropic polymer stress has increased enough to replace the missing entropic contribution. We argue instead that the trapped polymer conformations continue, microstate by microstate, to exert exactly the same mechanical force as they did when exploring the equilibrium ensemble. Since the thermodynamic tension in a chain is also the time-average of a thermally fluctuating mechanical tension, a `purely' entropic polymer stress can indeed be frozen in on vitrification. 
(However, if the sample is deformed again subsequently, this part of the stress has no reason to evolve as predicted by entropic elasticity, since polymer conformations are trapped with the `wrong' probability distribution for the new conditions.)  

A viewpoint that frozen polymers cannot exert an entropic stress would render our strain-hardening mechanism highly problematic in any system where the crinkle factor mechanism was not operative, so that the polymer stress is purely entropic throughout. (This might arguably include the case of shear deformation, considered next.) For in the strain-hardened state as $t\to\infty$, it is only because the polymers carry the stress that the solvent can remain frozen. Clearly if solvent freezing causes the polymer stress itself to disappear, then we have a paradox, and something else has to happen, such as a solvent stress that asymptotically approaches $\Sigma^s_Y$ from above as $t\to\infty$. The paradox is avoided if we are correct that an entropic polymer stress can persist within a frozen solvent. Note that closely analagous arguments concerning the persistence of entropic contributions to the chemical potential in frozen systems have been confirmed in both experiment \cite{Millner} and simulation \cite{Speedy}.

\section{Shear Deformation: Step Stress}
\label{shear}
Experimentally a tensile loading test as presented in \cite{Ediger} is a standard procedure for polymer glasses. The geometry is quite simple initially (a cylinder clamped at both ends) but the fact that the cross-section narrows with time means that the actual stress deviates from the engineering stress in the material and is not constant in time. This situation is made worse by `necking' which is commonplace in such tests and was ignored in the discussion above (but is briefly addressed within our model in \cite{PRL}).

From a theory or modeling perspective, it is interesting to ask what happens when a similar application and then removal of a constant load is made in a simple shear geometry. From the physics that was described in Section \ref{load} based on the precepts of standard dumb-bell elasticity, one expects in principle similar type of strain hardening to arise. The melting of the solvent at shear stresses above its yield stress, the gradual transfer of stress from solvent to polymer, and the subsequent revitrification of the solvent should all happen in a similar way.

\begin{figure}
\begin{center}
\includegraphics[width=76mm]{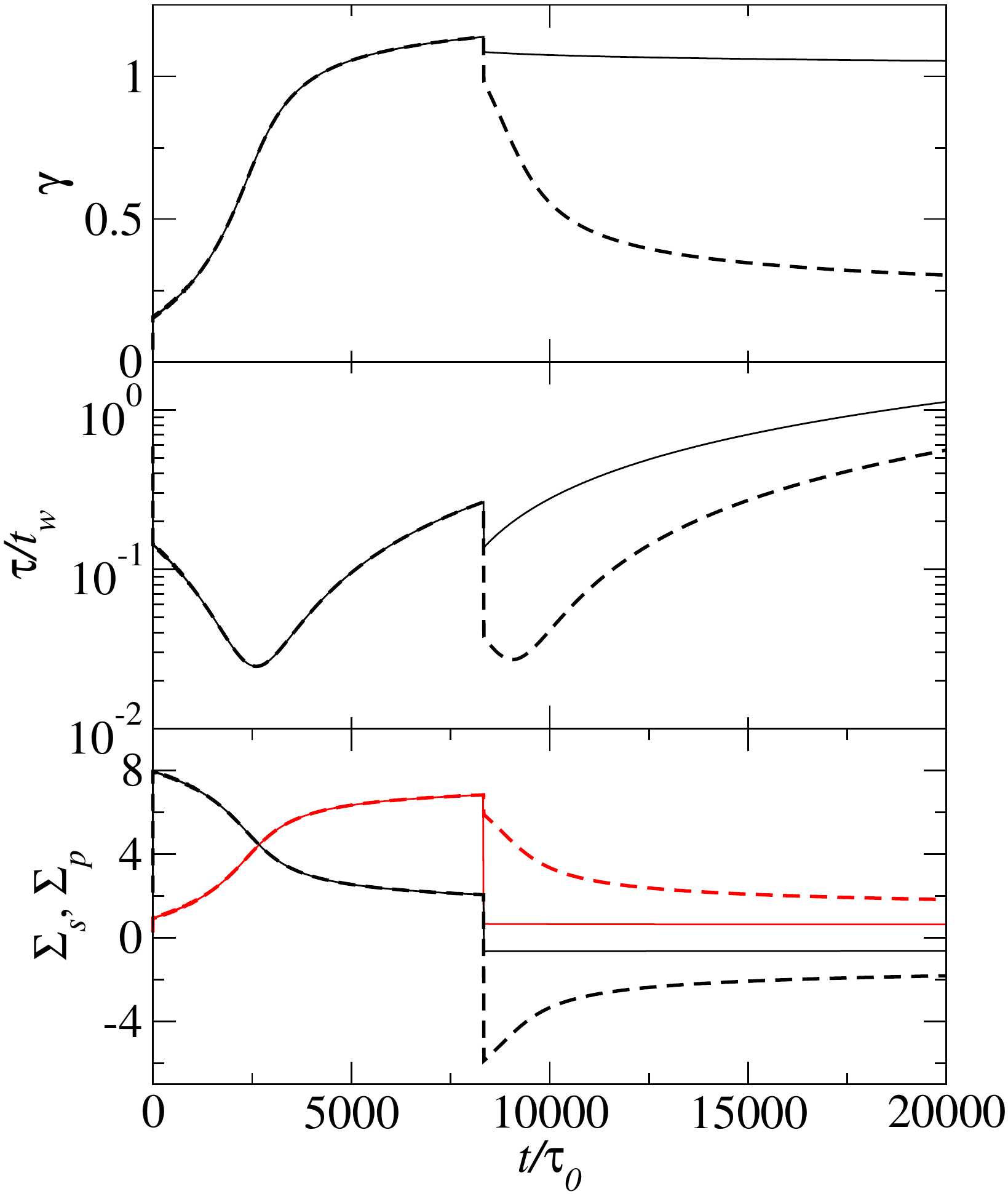}
\caption{ (Color online) Solid curves: shear strain $\gamma$, reduced relaxation time $\tau(t)/t_w$ 
and shear
stresses $\Sigma^{p,s}=G^{p,s}\sigma_{xy}^{p,s}$ of the
polymer (p) and solvent (s) during shear loading. Parameters as in Fig.\ref{fig_one}, with a matched ratio of shear stress to solvent yield stress.
(The curve for $\Sigma^p$, in red, initially lies below $\Sigma^s$ but crosses it during strain hardening.) 
The unload results for
the basic model ($\theta=1$) is shown dashed; the solid curve
after unload has $\theta = 0.1$.}
\label{fig_three}
\end{center}
\end{figure}

\begin{figure}
\begin{center}
\includegraphics[width=76mm]{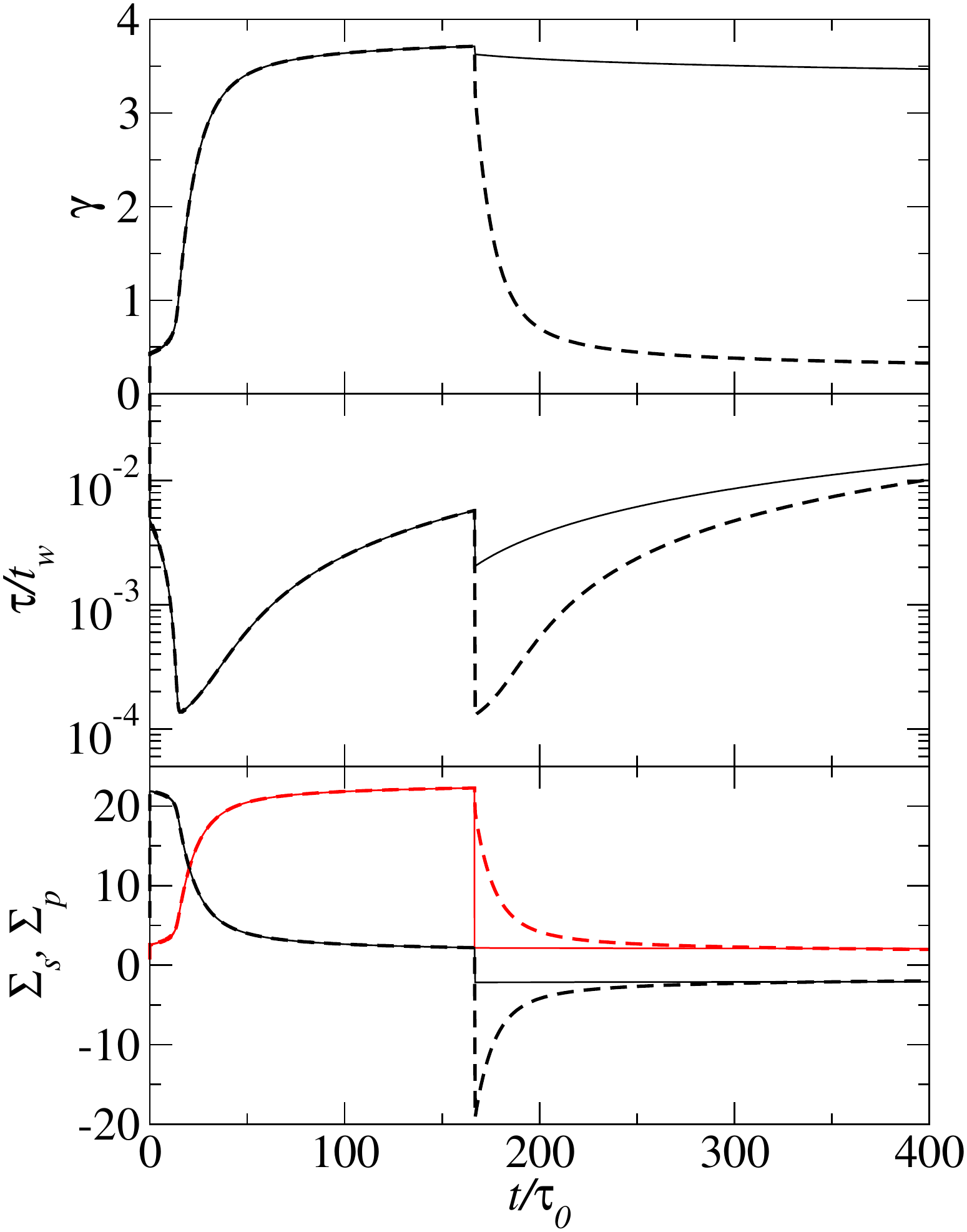}
\caption{ (Color online) As in Fig.~\ref{fig_three} but with shear stress increased by a factor 2.76. (This roughly matches the depth of the minimum in $\tau(t)$ to the elongational data of Fig.~\ref{fig_one}.) }
\label{fig_four}
\end{center}
\end{figure}

We therefore plot in Fig.~\ref{fig_three} the predicted response from our model for a shear load and unload protocol. This uses the same material parameters as in the fit to the experiments of \cite{Ediger} and is at a matched value of the engineering stress and the unload time to allow direct comparison with the elongational case as presented in Fig.~\ref{fig_one}. As anticipated above we find broadly similar behavior although the stress dip is far less pronounced. This latter effect may be due to the reduced strength of feedback during the initial shear-melting phase: recall that an additional geometric feedback (strain increasing the stress due to cross section change) is present only in elongation. Perhaps for similar reasons the dip, which is again synchronous with the stress transfer from solvent to polymer, appears somewhat later in shear than in elongation. In the baseline model with $\theta = 1$ (as might describe purely entropic elasticity) the second dip in $\tau(t)$ immediately after unloading is also less pronounced than in elongation. Introducing a crinkle factor $\theta = 0.1$ has similar effects to what happens in elongation, making this second dip still less pronounced and, perhaps  more importantly, drastically reducing the strain recovery achieved after unloading. Fig.~\ref{fig_four} shows the corresponding data for a substantially higher stress (six times the yield value). This is the value needed for the initial dip to match that seen in elongation (a factor $10^{-3.3}$ in $\tau/\tau(0)$); unsurprisingly it now happens very much earlier since the deformation builds far more quickly at this high stress. In this run we also reduced the unload time $t_u$ to roughly match $\tau(t_u)/\tau(0)$ at the point of unload.

To ensure comparability with the elongational results presented in Fig.~\ref{fig_one} and \cite{PRL}, the results of Figs.\ref{fig_three},\ref{fig_four} have been calculated under conditions where homogeneity of the flow is imposed. (The numerical protocols are also as detailed in \cite{PRL}.) An interesting question is whether the step stress loading protocol can create strain inhomogeneities such as shear bands (layers of large and small shear strain coexisting at equal stress). 
A theoretical criterion for shear banding in step stress has recently been derived; this requires that the underlying homogeneous creep curve $\gamma(t)$, differentiated to give $\dot\gamma(t)$, obeys $\partial^2_t\dot\gamma/\partial_t\dot\gamma >0$ \cite{TSB}.
We have explored this question numerically using the same methods as explained in Section \ref{startup} below for shear startup. We found that 
including noise in the initial condition can create persistent strain inhomogeneities throughout the loading and unloading phase. However, although these do not decay monotonically in time, they remain extremely small throughout. For example, 
with parameter values chosen as in Fig.~\ref{fig_four}, a fractional variation in strain rate, initially of order $10^{-3}$, decayed rapidly before rising slowly again to a peak whose height was itself less than $10^{-3}$. (The height of this peak was found to be linear in the initial noise amplitude.) The corresponding strain inhomogeneities remain smaller still for the more modest stress applied in Fig.~\ref{fig_three}. Moreover, we found that 
increasing the modulus, by curtailing the growth in shear rate through the strain hardening effect, suppresses strain inhomogeneity in step-stress relative to cases where $G^p$ is small but finite.
 
All this contrasts somewhat with the case of elongation where necking (ignored above but addressed in \cite{PRL}) is a ubiquitous phenomenon in polymer glasses. It stems mainly from the feedback between cross section and local stress (so that a region of narrowing in a cylindrical sample then narrows further).
It also contrasts with expectations based on recent work in startup of steady shear (see Section \ref{startup}), which does give shear banding, albeit transient, in fluidity models without polymer \cite{RobynPRL}.

Before addressing the case of startup (Section \ref{startup}) we return to the crinkle factor $\theta$. The reasoning for $\theta < 1$ given in Section \ref{more} was based on the picture of locally stretched chain segments creating high viscous dissipation \cite{LarsonKink,Hinch}; this picture was originally developed in the specific context of elongational flows. Shear flow of course also has an elongational axis, but for large strains this is not aligned with the polymer orientation (the latter approaches the flow direction while the elongational axis remains at 45 degrees). For this reason, within the picture we have developed, one can perhaps expect a much weaker buildup of non-entropic stresses and hence a much smaller drop in polymer modulus on unload. Indeed, taken at face value our arguments suggest a value of $\theta$ close to unity, which in turn requires that the polymer modulus relevant to shear flow is the entropic one, and hence approximately ten times smaller than that used to fit the elongational loading data \cite{foot}.

If this grossly oversimplified picture were correct then one could expect very significant differences in the experimental curves, not only for $\tau(t)$ but also for strain, between elongational and shear loading. Crudely speaking (i.e., ignoring the required shift in $G^p$) the elongational experiment would follow the {\em solid} line in Fig.~\ref{fig_one}, as indeed it nearly does, but the shear protocol would follow the {\em dotted} line in Fig.~\ref{fig_three}. The most important macroscopic consequence is that in shear the strain recovery on unload would be much larger, because the polymer loading on the solvent, which is nearly the same in magnitude as the initially applied load, is always enough to re-melt the solvent (unless the initial load was only slightly above the solvent yield stress).  

To counter this argument, it has recently been suggested that strain hardening in polymer glasses is attributable not to the viscous stress on a kinked filament but nonetheless to the `pulling tight' of a small fraction of chain strands trapped between entanglements \cite{TomPaper}. This is quite similar in spirit to the mechanism discussed above, and could also lead to small $\theta$, but could be somewhat less dependent on the deformation geometry.  
To help understand these issues further, it would be extremely interesting to see experimental comparisons between tensile and shear loading on similarly prepared samples at comparable stress.

\section{Shear Startup}
\label{startup}
To check that our simple model is not limited to the step-stress case but also behaves reasonably in strain-controlled flows, we calculated in \cite{PRL} (as supplemental material) the stress responses for startup of steady elongation and compression. These show a stress overshoot, whose height varies as $\ln(\dot\varepsilon t_w)$, which is similar to the behavior found (under shear) in simple aging fluids \cite{SGR}, and also broadly accords with reports in the polymer glass literature (see e.g. \cite{EGP,EGP3}.

Here we extend this further to address the startup of steady shear which is a widespread experimental protocol in the polymer rheology literature generally. The physics of the stress overshoot involves a transient elastic deformation that exceeds the recoverable strain in the steady state that is reached eventually. This is common in systems with strong shear thinning whose initial elastic response involves structures, or polymer conformations, that can be reached at modest strains from the initial state but cannot be sustained thereafter.

\begin{figure}
\begin{center}
\includegraphics[width=76mm]{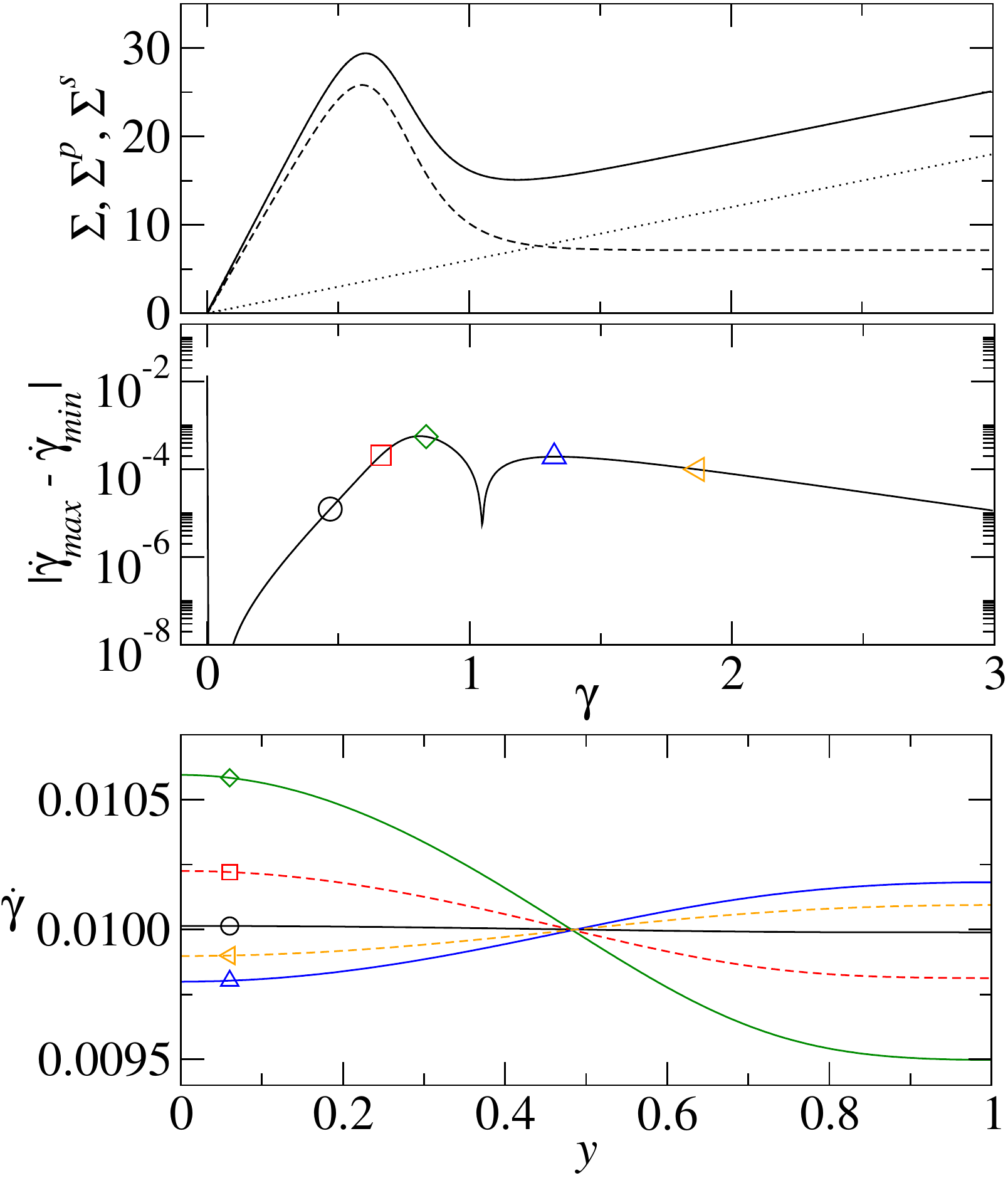}
\caption{ (Color online) Startup of steady shear at applied strain rate $\dot\gamma = 10^{-2}$ for a system with the parameters of Fig.~\ref{fig_one}. Upper panel: Total shear stress (solid) and the polymer (dotted) and solvent (dashed) contributions when the flow is imposed to be uniform. 
The additional dotted points barely distinguishable from the solid curve show the total stress after shear banding is allowed for. 
Middle panel: the `degree of banding' (found by subtracting the smallest from the largest shear rate present at any time) for a 2D run with heterogeneity allowed. (A small diffusivity was added to the governing equations for all stress components and for $\tau$, and the system was initialized with a small spatially varying noise; see \cite{footmethods}.) Lower panel: snapshots of the strain rate as a function of position $y$ in the flow gradient direction with symbols identifying strain values as in the middle panel.}
\label{fig_five}
\end{center}
\end{figure}

Figure \ref{fig_five} shows the polymer, solvent and total contributions to the shear stress during startup for parameters as in Fig.~\ref{fig_three} at an imposed strain rate $\dot\gamma = 10^{-2}$. These curves are calculated numerically on the assumption that the flow remains homogeneous. 

However, recently it has become clear that overshoots of this kind, which are particularly common in aging systems, often cause instabilities leading to transient shear banding (layers of material with different $\dot\gamma$ at a common stress) \cite{Manning,RobynPRL,TSB}. These instabilities typically occur within the region where stress is decreasing and strain increasing, which can crudely be viewed a negative differential elastic constant (although this picture is certainly oversimplified \cite{TSB}). We therefore repeated the simulations allowing the flow to become nonuniform and indeed find transient strain rate variations with position that rise to a maximum near the stress peak and then decay only very slowly; these results are also in Fig.~\ref{fig_five} as is the total stress arising when the nonuniformity is taken into account. (The numerical methods are summarized in \cite{footmethods}.) 

We should emphasize that for these parameters (chosen to match the elongation experiments of \cite{Ediger}), the shear banding is a very weak effect. Although the system is transiently unstable to the formation of inhomogeneities in the velocity gradient, these barely deserve to be called shear bands. Indeed the maximum deviation in shear rate plotted in Fig.~\ref{fig_five} depends directly on the initial noise used to excite the perturbation. For the data shown the velocity inhomogeneity in the initial state (which decays rapidly in the first few timesteps) was in fact larger than the maximum recorded subsequently.

\begin{figure}
\begin{center}
\includegraphics[width=86mm]{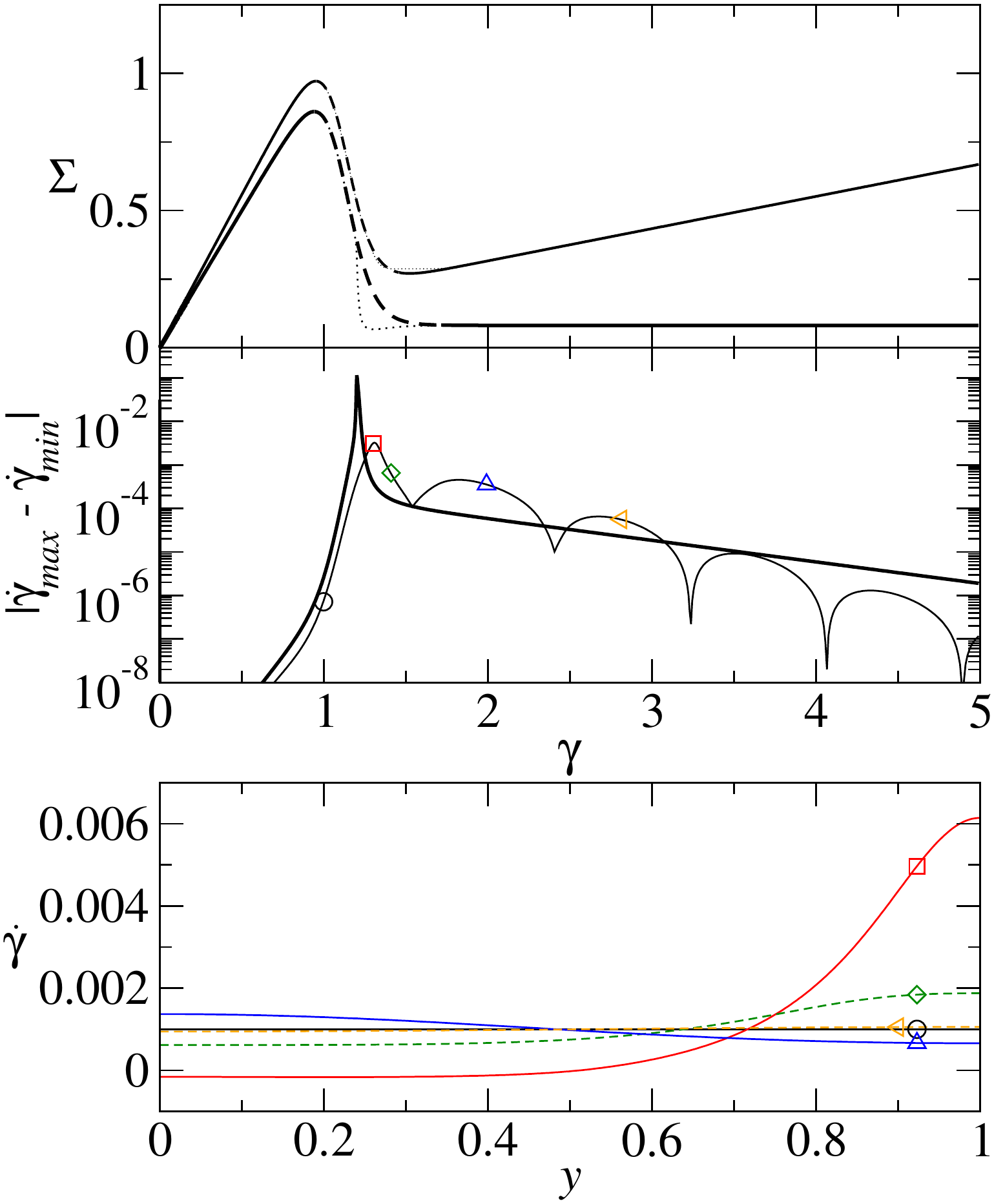}
\caption{(Color online) Upper panel: shear stress as a function of time in shear startup for the fludity model without (lower curve) and with polymer. ($G^p/G^s = 0,1/8.5$ respectively.) Dashed regions: transient instability as found by linear stability analysis \cite{footmethods}. Dotted curve is total stress allowing for inhomogeneous flow. Middle panel: the resulting absolute value of shear rate variations with polymer (multiply bumped curve with added symbols) and without (singly cusped curve) \cite{footmethods}. 
Lower panel: snapshots of strain rate profile with polymer present, at strain points identified by the symbols as in middle panel. Note the strong strain inhomogeneity (shear banding) at strains just beyond the stress overshoot.
Parameter values $G_s,\tau_0 = 1$,  $\mu = 12.5$, $t_w = 10^8$, $\dot\gamma = 10^{-3}$. Other parameters as in Fig.~\ref{fig_one} and \cite{footmethods}.
}
\label{fig_six}
\end{center}
\end{figure}

However, the banding does become robust and reproducible at much larger values of $t_w/\tau_0$ where we recall that $t_w =\tau(0^-)$ is the sample age at the initiation of shear. 
For one such case ($t_w/\tau_0 = 10^8$), this is demonstrated in Fig.~\ref{fig_six}, which also shows a comparison between our polymer model and a simple fluidity model obtained by setting $G^p$ to zero. A linear stability analysis \cite{footmethods} shows that the flow becomes unstable, as expected, in the decreasing part of the overshoot but that the additional polymer stress, which has no overshoot, exerts a slight stabilizing role. To confirm this we studied the real part of the largest eigenvalue in the stability analysis as well as the resulting strain variation amplitude; both decrease monotonically with $G^p/G^s$\if{ as shown in Fig.~\ref{fig_seven}}\fi. Despite this, the presence of polymer does add significant complication to the deviations from uniformity that actually arise, whose magnitude now shows an oscillation in time that was not present in the simple fluidity model (Fig.~\ref{fig_six}). 

In shear startup, the general effect of increasing the polymer modulus is to decrease the degree of transient banding observed. This is because the polymeric stress contribution increases monotonically with strain (there is negligible polymer relaxation) which decreases the severity of the stress decline after the overshoot in the total stress. Because transient shear banding is triggered by the declining stress \cite{TSB}, it is mitigated by the polymeric contribution.

\if{
\begin{figure}
\begin{center}
\includegraphics[width=86mm]{JCP_FIG_7.pdf}
\caption{Upper panel: A measure of instability defined as the time integral of a function equal to the real part of the largest eigenvalue when this is positive and zero otherwise. This is plotted against $G^p/G^s$ showing the stabilizing effect of polymers on the shear-startup instability of the glass. Solid line: $t_w/\tau_0 = 10^8$; dashed line $t_w/\tau_0 = 10^{10}$. Bottom: the maximum value of the strength of the shear banding. (Other parameters as in Fig.~\ref{fig_six}.)}
\label{fig_seven}
\end{center}
\end{figure}
}\fi
\section{Conclusion}

In this paper we have developed a simple approach, initiated in \cite{PRL}, to the modeling of polymer glasses. This builds on recent models of rheological aging and rejuvenation in simple glassy fluids, coupled to a minimal model of polymers at dumb-bell level. The presence of single polymer mode rather than a Rouse spectrum means that we cannot address the non-exponential form of local relaxations that was reported in \cite{Ediger}; however it would be simple in principle, if cumbersome numerically, to add additional modes to our picture. An alternative route to nonexponential relaxation is through the solvent sector where the fluidity model similarly replaces a complicated relaxation lineshape (ubiquitous in simple glasses close to $T_g$) with a single relaxation time. Improvements are probably achievable without leaving the fluidity framework but only at the cost of additional parameter fitting.

That issue apart, our model can explain much of what happens experimentally \cite{Ediger} when a polymer glass is subjected to elongational load. The unloading behavior is more problematic, but consistent with a plausible modification of the same model which invokes on unload a reduction in polymer modulus via a `crinkle factor' $\theta < 1$. This crudely allows for the presence of non-elastic polymer stresses that are known to arise whenever $\dot\varepsilon\tau^p$ is large \cite{LarsonKink,Hinch,CCH}. We tentatively associate these stresses with the observed uplift in the polymer modulus $G^p$ beyond the entropic value $G^e$ that would arise for non-glassy polymers, so that $\theta \simeq G^e/G^p$. However, this relies on physical reasoning that was so far developed in the literature primarily to address elongational deformation, not shear.

Partly with this in mind, and in the hope of stimulating new experiments, we have presented here analogous calculations for a shear loading geometry. On loading, these predict very similar behavior for the strain and the segmental relaxation time $\tau(t)$ although the characteristic `dip' in the latter quantity is somewhat less pronounced. The unloading behavior is again similar if the same crinkle factor is used. However, it is at least conceivable that $\theta$ in this non-elongational case should be close to unity. If so, the model predicts a very different unloading response in which the polymer stress is enough to remelt the solvent, causing a drastically larger strain recovery than seen in elongation. Such an interpretation would also imply a much smaller non-entropic uplift in polymer modulus for shear flow than for elongation. 

We have also presented new results for startup shear flows and shown that, in common with other shear-thinning glassy materials, transient flow inhomogeneities can be expected in this case. However these mainly stem from the glass physics of the solvent itself, and in fact the polymers have a mild stabilizing effect on such instabilities. In the polymer glass context they may therefore prove inessential, although their possible presence should certainly be borne in mind when interpreting experiments.  

Our work suggests that an accurate representation of aging and rejuvenation physics may form a key part of any more comprehensive theory of polymer glass rheology. It also suggests that insights gleaned from studying the rheology of simple glasses will form a useful input into such a theory. As shown in this paper, some of these insights (such as simple aging and shear rejuvenation) can be embodied in models that are simple enough to be coupled with standard, if simplified, polymer rheology models such the dumb-bell representation. In general it seems a good strategy for future work to improve the glass and polymer sectors in tandem, rather than let one become much more sophisticated than the other. More generally our hope is that a more comprehensive account of polymer glasses can be achieved by judiciously combining existing types of nonlinear theory, describing non-glassy polymers and simple glasses respectively. 

Of course, it may be that polymer glasses involve `new physics' that is not present in either such description, but it may also be that many of their properties emerge simply from the combination. Strain hardening, according to our interpretation, is a case in point. The task of combining polymer and glass theories in this way is not necessarily a simple one and could raise various issues of principle. We have discussed one of these in this paper, namely the question of what is meant by entropic elasticity in a system whose conformational state is frozen. We believe we have the right answer to this question (the entropic force is maintained but its response to further deformation is altered), which mirrors similar discussions of phase equilibria involving nonergodic materials that date back over 80 years \cite{Millner,Speedy,Speedy2}. This type of question might remain a technical nicety if the predominant polymer stress in glasses is never entropic \cite{Robbins,hoy}, but comes to the fore if the opposite is ever true.

\subsection*{Acknowledgements} 
MEC is funded by a Royal Society Research Professorship. This work was funded in part by EPSRC EP/E030173 and EPSRC/E5336X/1.
The research leading to these results has received funding to SMF from the European Research Council under the European Union's Seventh Framework Programme (FP7/2007-13)/ERC Grant agreement no. 279365.
RGL is partially supported from NSF under grant DMR 0906587. Any opinions, findings, and conclusions or recommendations expressed in this material are those of the authors and do not necessarily reflect the views of the National Science Foundation (NSF).


\begin{thebibliography}{}
\bibitem{footcon} The resulting model is clearly not quite consistent because the reduced modulus should then also control the early stages of the response after the initial loading before chains become locally stretched. 
Therefore, at some stage during the transfer of stress from solvent to polymer visible in Fig.~\ref{fig_one}, the modulus should smoothly rise from $\theta G^p$ to $G^p$ to reflect the onset of local inextensibility. However, we do not expect this drift in modulus to give qualitative changes in the predictions for the loading part of the experiment: what mainly matters there is the modulus within the strain-hardened regime.
\bibitem{foot} Note that this viewpoint would be hard to sustain if one believes that the polymer entropic stress disappears on vitrification of the solvent as mooted in Section \ref{mmore}.
\end{thebibliography}
\end{document}